\documentstyle[11pt,aasms4,rotating]{article}
\def\msun{{\,M_\odot}}

\def\simlt{\lower.5ex\hbox{$\; \buildrel < \over \sim \;$}}
\def\simgt{\lower.5ex\hbox{$\; \buildrel > \over \sim \;$}}

\def\gcm3{{\rm\,g\,cm^{-3}}}
\def\ncm3{{\rm\,cm^{-3}}}

\def\>{$>$}
\def\<{$<$}

\lefthead{Melia et al.}
\righthead{}

\begin{document}
\centerline{Submitted to the Editor of the Astrophysical Journal Letters}
\vskip 0.5in
\title{\bf New Constraints on the Nature of Radio \\ Emission in Sagittarius A*}

\author{Siming Liu\altaffilmark{1} and Fulvio Melia$^{1,2,3}$}

\affil{$^1$Physics Department, The University of Arizona, Tucson, AZ 85721}
\affil{$^2$Steward Observatory, The University of Arizona, Tucson, AZ 85721}


\altaffiltext{3}{Sir Thomas Lyle Fellow and Miegunyah Fellow.}


\begin{abstract}

The mm to sub-mm spectrum of Sgr A* at the Galactic center, as well as its
polarization characteristics, are consistent with the inner 10 Schwarzschild radii
of a tight Keplerian emitting region of hot, magnetized, orbiting gas.  This plasma
may also be the source (through self-Comptonization) of the X-rays detected by {\it
Chandra}.  It has long been suspected that the circularization region between the
quasi-spherical infall at large radii, and this inner zone, is responsible for
producing the rest of Sgr A*'s spectrum. In this paper, we report the results of a
detailed study of this region, with several important conclusions that will be
highly relevant to upcoming coordinated multi-wavelength observations. First, the
combination of existing cm and X-ray data preclude the possibility of producing the
observed strong 1.36 GHz radio flux via thermal synchrotron within a bounded flow.
If Sgr A*'s radio spectrum is produced by accreting gas, it appears that a
non-thermal particle distribution is a necessity. This may not be surprising, given
that the energy associated with the radial motion is probably dissipated by shocks
before the gas circularizes, which can produce the required power-law distribution.
Second, if this is the correct picture for how Sgr A*'s spectrum is produced, it
appears that the {\it Chandra}-detected X-rays may originate {\it either} from
self-Comptonization in the inner Keplerian region, or from optically-thin
nonthermal synchrotron emission in the much larger, circularization zone, extending
up to 500 Schwarzschild radii or more.  This is a question that should be answered
by upcoming broadband observations, since the mm-bump and X-rays are strongly
correlated in the former case, whereas the X-rays are strongly correlated to the
cm-radio flux in the latter. In addition, X-rays produced in the circularized gas
could show periodic or quasi-periodic variations, but not those produced via
nonthermal synchrotron emission much farther out.

\end{abstract}


\keywords{accretion---black hole physics---Galaxy: 
center---hydrodynamics---magnetic 
fields: dynamo---radiation mechanisms: nonthermal}


%

\section{Introduction}

A concentration of $2.6\times 10^6\,\msun$ dominates the gravitational potential
within $0.015$ pc of the Galactic center (Genzel et al. 1996; Ghez et al.  1998).
The extremely compact, nonthermal radio source, Sgr A* (Krichbaum et al.  1993;  Backer et al.
1993; Krichbaum et al. 1998; Lo et al. 1993), is thought to coincide with this black
hole candidate. Its radio spectrum
from cm to mm wavelengths is roughly a power-law, $S_{\nu}\propto\nu^a$, with $a\sim
0.19-0.34$. The sharp cut-off at the low frequency end ($\sim 0.5$ GHz) appears to be
due to free-free absorption in the ionized gas along the line-of-sight (Davies et al.  
1976). In the mm region, a spectral bump (Zylka et al. 1992; Zylka et al.  1995) has
been confirmed by simultaneous $20$ cm to $1$ mm radio observations (Falcke, et al.
1998), suggesting that a distinct emission component surrounds the event horizon,
since the highest frequencies appear to correspond to the smallest spatial scales
(Melia, Jokipii \& Narayanan 1992; Melia 1992, 1994; Nayaran, Yi \& Mahadevan 1995;
Falcke et al.  1998; Coker \& Melia 2000). The spectrum turns over in the sub-mm
range.

Several different scenarios have been introduced (see Melia \&
Falcke 2001) to account for Sagittarius A*'s spectrum, ranging from jet emission
(Falcke \& Markoff 2000) to heavily-advected, large-scale accretion disks (Narayan,
Yi \& Mahadevan 1995), a picture that in recent years has spawned several other
versions, including disks with mass-loss (Blandford \& Begelman 1999), and those with
significant convection (Narayan, Igumenshchev \& Abramowicz 2000).  The approach 
followed by our group (see Melia 1992) has been based on the idea that Sagittarius
A* may be accreting a low specific angular momentum plasma, which radiates
inefficiently because its magnetic field is very low (Kowalenko \& Melia
1999; Coker \& Melia 2000), or is otherwise weakly luminous because the
rate at which gas actually reaches the event horizon is smaller than one
would infer based on simple Bondi-Hoyle estimates (see, e.g., Quataert
\& Gruzinov 2000; Melia, Liu \& Coker 2000).  Large scale 3D
hydrodynamic simulations (e.g., Coker \& Melia 1997) do show (not surprisingly) that
in the absence of an anomalously high viscosity near the Bondi-Hoyle capture radius,
the accreted gas has a small, though variable, specific angular momentum $\lambda\ c\
r_S$, where $\lambda < 40$ and $r_S\equiv 2GM/c^2$ is the Schwarzschild
radius. These values of $\lambda$ may be consistent with the specific angular
momentum associated either with the thermal motion of the gas or with the orbital
velocities.

With such a small value of $\lambda$, the infalling gas does not circularize
until it reaches a radius $\sim 2\lambda^2\,r_S$, and depending on the
physics of magnetic field annihilation, or generation (see, e.g., Hawley et al. 
1996), may not become an efficient radiator until it falls to within tens of
Schwarzschild radii of the event horizon. Recent radio polarization 
measurements (Bower et al. 1999; Aitken et al. 2000) seem to support the 
spectral signature expected for such a low angular momentum gas.  In
Melia et al. (2000, 2001), we demonstrated that a small Keplerian structure, 
with a magnetic field dominated by its azimuthal component inferred from
earlier magnetohydrodynamic simulations (Brandenburg et al. 1995; Hawley 
et al. 1996), not only produces the mm bump, but also accounts for these 
polarization characteristics, especially the $\sim 90^\circ$ flip of the 
polarization angle near the mm peak (Bromley et al. 2001). However, this 
inner Keplerian region does not produce the longer-wavelength emission.
The radio photons from Sagittarius A* must be produced at larger radii, 
in the circularization region where the infalling gas evolves from 
quasi-spherical accretion toward a settled Keplerian configuration at smaller radii.
In this {\it Letter}, we report the results of a detailed analysis of this 
circularization region, and demonstrate how the latest combination of cm, 
and X-ray data already constrains heavily the nature of its emissivity, 
and at the same time provides a telling discriminant between two principal 
mechanisms for producing the high-energy emission in this source. 

\section{Physics of the Circularization Region}

Having identified the essential elements of the inner Keplerian structure that
correctly account for the mm and sub-mm spectrum and polarization of Sagittarius A*
(Melia et al. 2000, 2001), we here take these physical quantities as
inner boundary conditions for the circularization region. The strength of the
magnetic field in the inner region is calculated based on the magnetohydrodynamic
simulations of Brandenburg et al. (1995) and Hawley et al. (1996).  These results are
conveniently expressed in terms of two parameters, the ratio of the stress to the
magnetic energy density, $\beta_\nu$, and the ratio of the magnetic energy density to
the thermal pressure, $\beta_p$. 

Because the specific angular momentum of the infalling gas is assumed to be small,
the accretion at large radii can be approximated as quasi-spherical.  Thus, to match
the Keplerian flow at small radii (typically within the inner $10r_S$), the gas
kinetic energy associated with the radial motion in the quasi-spherical region must
be dissipated into thermal or turbulent energy. 
The centimeter radio emission is presumably produced as a result of this
dissipation.  We will assume that the circularization region has a flattened shape,
and that its vertical structure is determined by balancing the thermal pressure
gradient with the gravitational force in that direction. At the same time, we also
assume that $\lambda$ is conserved in the circularization
region, with the justification that most of it is dissipated once the plasma
circularizes. This is clearly only an approximation, and in reality, we expect there
to be some angular momentum dissipation as the gas approaches the Keplerian region.  
Thus, specifying the outer radius of the latter to be $r_o$, we have that the
azimuthal velocity is given as $v_\phi(r) = \sqrt{G\ M\ r_o}/r$.

For the temperature profile, we assume that a fixed fraction of the gravitational
potential energy is dissipated into thermal energy and that this fraction is fixed by
the value of the gas temperature $T(r_o)$ at $r_o$.  In that case, $T(r)= (G\ M/r-
0.5\ v_\phi^2)f_t/\alpha R_g$, where $f_t = 2.0\ T(r_o)\ \alpha\ R_g\ r_o/G\ M$,
$R_g$ is the gas constant, and $\alpha\sim3.0-4.5$ is only weakly dependent on the gas
temperature (Coker \& Melia 2000). The exact value of $f_t$ is determined by the
partition of energy in the turbulent accretion flow, which depends on how dissipation
drives the large scale turbulent motion, how effectively the energy is transported
to small scales and converted into thermal energy, and then radiation. 
Unfortunately, all these processes are not well understood.
Based on the equipartition argument, we know $f_t$ is smaller than one.
Together with the above expression for $v_\phi(r)$, this then fixes the kinetic
viscosity $\mu = \beta_\mu\ H\ c_s$, where $H=\sqrt{2R_g\ T\ r^3/G\ M}$ is the scale
height of the circularization region, $c_s=\sqrt{2R_g\ T}$ is the isothermal sound
speed and $\beta_\mu=(2/3)\ \beta_p\ \beta_\nu$.  It follows that $\mu(r)\ \rho(r)
H(r)$ is a constant in this region, and this uniquely determines the radial profile.  
When fitting Sgr A*'s radio spectrum below, we shall find that only the
circularization region's outer radius remains as an adjustable parameter.

One of the key results we report here is that the radio emission from Sagittarius
A* cannot be produced via {\it thermal} synchrotron processes within a bounded
accretion flow.  This is rather easy to show, and is based on the fact that with
the temperature limited from above by its virial value, the required radio emission
can be produced by thermal synchrotron only with a strong magnetic field, whose
energy density must be 15 times larger than its equipartition value to avoid
excessive X-ray emission via thermal bremsstrahlung process. Attempts to produce
the observed radio spectrum with a more reasonable equipartition magnetic field
result in high electron number densities, which in turn violate the observed X-ray
flux (Baganoff et al. 2001).

To see this in detail, we have for a fully ionized bounded gas that $T(r)< 1.8\times
10^{12} r_S/r$ K, where $r_S=7.7\times 10^{11}$ cm for the Schwarzschild radius of
Sgr A*.  The thermal emission from the circularization region is black-body limited,
so that to produce a $0.53$ Jy radio flux density at $1.36$ GHz with a thermal source
at the Galactic center, we need $T(r)\ (r/r_S)^2>3.4\times 10^{15}$ K, where $r$ is a
characteristic radius in the emitter. Combining this with the virial temperature
limit, we have $r>1900\ r_S$. The corresponding temperature limit is $10^9$ K. Now,
to suppress the bremsstrahlung X-ray emission from a gas with temperature $10^9$ K
below the observed value, the ratio of the cyclo-synchrotron emissivity at $1.36$ GHz
to the bremsstrahlung emissivity at $10^{18}$ Hz must be larger than the ratio of
the observed flux densities, since part of the radio emission is self-absorbed. We
can estimate the bremsstrahlung emissivity at $10^{18}$ Hz as $\epsilon_b=6.0\times
10^{-42} n^2$ ergs s$^{-1}$ Hz$^{-1}$ cm$^{-3}$ (Melia 1992), and the
cyclo-synchrotron emissivity at $1.36$ GHz as $\epsilon_s=4.2\times 10^{-17}n\ M(x)$,
where $M(x)=0.1746\ e^{-1.8899\ x^{1/3}}/x^{1/6}$ when $x$ is much bigger than 1 and
$x=1.1\times 10^4/B$ (Mahadevan et al. 1996).  If the magnetic field is in
equipartition (a conservative upper limit), then $B^2 = 1.0\times 10^{-5} n$, and
from $\epsilon_b<2.8\times 10^{-8}\epsilon_s$, we get that $B$ must be larger than
$8.1$ G.  However, the electron number density is then $n>6.6\times 10^6$ cm$^{-3}$,
and for a source size of $1,900 r_S$ at the Galactic center, the X-ray flux density
at $10^{18}$ Hz produced via thermal bremsstrahlung will be $3.9\times 10^{-5}$ Jy,
which exceeds the observed limit.

We conclude from this that the {\it thermal} radiative efficiency of a bounded
accretion flow is not sufficient to account for the cm-wavelength spectral
component of Sgr A*. We ask, therefore, whether this long wavelength emission may
be due to nonthermal particles produced by the gas in transition.  The possible
contribution of a nonthermal particle component in the accretion flow of Sgr A* was
introduced by Markoff, Melia, \& Sarcevic (1997), primarily to explore whether the
EGRET $\gamma$-ray source 2EG J1746-2852 could be due to interactions among the
decay products resulting from relativistic $p-p$ scatterings.  Within the context
of the ADAF model, Mahadevan (1998) introduced the idea of accounting for the cm
radio spectrum with a power-law electron distribution also produced via $p-p$
scatterings, but the predicted X-ray spectral index in this picture is not
consistent with that observed.  An alternative scenario, involving a static,
quasi-monoenergetic electron distribution which may be produced by magnetic field
reconnection, was considered by Duschl \& Lesch (1994).

But relativistic electrons can also be produced by direct shock acceleration, and
with our expectation that shocks may be present in the supersonic flow of the
turbulent circularization region, it is reasonable to assess the merits of such a
power-law distribution to account for Sgr A*'s radio spectrum, while at the same
time not violating (or perhaps even simultaneously explaining) its X-ray spectrum
as well. An important departure of this work from the earlier attempts at carrying
out this fit is the much lower accretion rate we are adopting here, which
significantly changes the interplay between the radio and X-ray emissivities. We
will see below, in fact, that depending on the efficiency of converting
gravitational energy to thermal energy, the X-rays can be produced either via
self-Comptonization in the Keplerian region or via synchrotron emission by the
power-law electrons in the circularization region.  This is clearly highly relevant
to future coordinated multi-wavelength observations, which should be able to
distinguish between these two possible manifestations of the high-energy radiation.

However, even a cursory inspection of the data will immediately show that the
nonthermal electrons must have a rather steep spectral index. Strong shock
acceleration usually produces a power-law with an index of $\sim 2.0$, much smaller
than our best fit value (see below). There are several reasons that can account for
this. First, it is well known that the shock compression ratio $p_c$, which
determines the spectral index according to $p=(p_c+2)/(p_c-1)$, can be reduced
significantly in a strongly magnetized plasma (Kirk et al. 2000; Ballard \& Heavens
1991). In our model, we have a magnetic field which is strong enough to decrease the
compression ratio well below $2.5$, which can therefore yield a power-law spectral
index larger than $3.0$. Second, the circularization region may be turbulent, with a
very irregular magnetic field.  Kirk et al. (1996, 1997) have demonstrated that
braided magnetic fields can increase the power-law spectral index even further, to
$(p_c+3.5)/(p_c-1)$.  Third, Duffy et al. (1995) noticed in their simulations that
the accelerated ions can lead to the creation of a weaker subshock and an upstream
precursor, whose effect also leads to a reduction of the compression ratio and a
steeper spectral index.  An additional factor that may be relevant here is that with
rapid cooling (we will see below that in some cases the nonthermal particles cool
well before reaching the inner Keplerian region) the steady-state distribution
function is characterized by a spectral index $p+1$ (Laan 1963; Markoff et al. 1999).
Even so, turbulent shock acceleration is not well understood, so we must necessarily
treat the spectral index and the number density of nonthermal particles as free
parameters.  For simplicity, we will assume that the former is the same throughout
the circularization region, and that the ratio of the power-law electron number
density to the total electron number density in the flow is constant. We shall see
that the combination of cm, mm, and X-ray data do not leave much flexibility in the
possible values of these quantities. In fact, our best fits (see Figs. 1 \&
2) require an approximate equipartition between the thermal and nonthermal particle
distributions in the circularization region. 

\section{Results and Discussion}

Our two best fit models are shown in Figures 1 (model 1) and 2 (model 2).
The flattened gas distribution has an inclination angle of $45^\circ$
to the line of sight, and the circularization region extends out to $1,000\ r_S$.  
The radio spectrum results from the superposition of power-law synchrotron emission
components at different radii, with the lowest frequency part produced in the large,
diluted outer segments, and the high frequency portion
resulting from emission in the smaller, dense inner zone. The overall spectrum is
relatively flat from $1$ GHz to $50$ GHz. From Figure 1, it is clear that the thermal
emission from this region is negligible by comparison. The radiation below $1$ GHz,
which is produced at radii beyond $500\ r_S$, is mostly absorbed by the ionized gas
along the line-of-sight (Davies et al. 1976). This type of fit to the radio spectrum
therefore requires a circularization region extending beyond $500\ r_S$, but the
upper limit to this radius is not well-defined.  To fit the observed spectrum below
$1$ GHz, we assume that the absorbing gas has a temperature of $10^4$ K, and we use
the free-free optical depth given by Walker et al. (2000).

Model 1 has a relatively low efficiency of converting gravitational to thermal
energy. The temperature in the accreting gas stays relatively low,
increasing from $\sim 2\times 10^8$ K at $1,000\ r_S$, to its maximum value of
$1\times 10^{11}$ K at the inner edge of the Keplerian structure. From Figure 1, it
is evident that the electrons are not energetic enough to up-scatter radio photons
into the X-ray band. So for this situation, the power-law electrons in the
circularization region are producing the X-rays directly via optically-thin 
synchrotron emission.  The predicted X-ray spectral index $a$ is therefore $(p-1)/2$, 
where $p$ (the particle spectral index) is
fixed by the relative strength of the radio to X-ray flux densities. Our best fit for
the radio and X-ray components requires a value $a = 1.05$, which falls within the
{\it Chandra}-measured range of $0.75$ to $3.0$ (Baganoff et al. 2001).  In this
case, we expect there to be strongly correlated variability between the cm-radio
emission and that at X-ray energies (since both are produced within the
circularization region), while the mm bump would not be as strongly correlated with
the X-rays.

Model 2 has a much higher efficiency of energy conversion, and almost all of the
dissipated gravitational energy goes into thermal energy. The inner Keplerian region
has a temperature of $1.0-3.0\times 10^{11}$ K, which is high enough to produce
strong sub-mm and infrared emission, and the electrons can 
up-scatter some of them into the X-ray band. In this case, we would expect to see a
strong correlated variability between the mm radio bump and the X-rays. However,
in order to suppress the X-ray emission in the circularization region, the power-law
electrons must have an index larger than $3.2$.

Whether or not self-Comptonization dominates the high-energy emission can be
determined with future coordinated multi-wavelength observations.  If the cm-radio
photons and the X-rays are produced by a single power-law, the particle spectral
index must then be $\approx 3.1$ (model 1). However, if the X-rays are due mostly
to self-Comptonization from the inner Keplerian region, this index must be larger
than $3.2$ (model 2).  An alternative prescription that can also avoid producing
too many X-rays and infrared radiation is to introduce a cutoff in the particle
energy. The required limit on the Lorentz factor is then $\gamma_e<100$, which may
result from particle acceleration due to magnetic reconnection or hydromagnetic
wave turbulence (Litvinenko, 2000). Finally, it can also be shown that the
electrons producing the optical emission have a lifetime shorter than $15$ minutes.
This is much shorter than the advection time scale of $20$ hours through the zone,
so the nonthermal particles advected into the Keplerian region contribute little to
the overall emission there.

{\bf Acknowledgments} 
This research was partially supported
by NASA under grants NAG5-8239 and NAG5-9205, and has made use of NASA's
Astrophysics Data System Abstract Service.  FM is very grateful
to the University of Melbourne for its support (through a
Miegunyah Fellowship). 

{}

%
%

\begin{figure}[thb]\label{fig1.ps}
{\begin{turn}{0}
\epsscale{0.8}
\centerline{\plotone{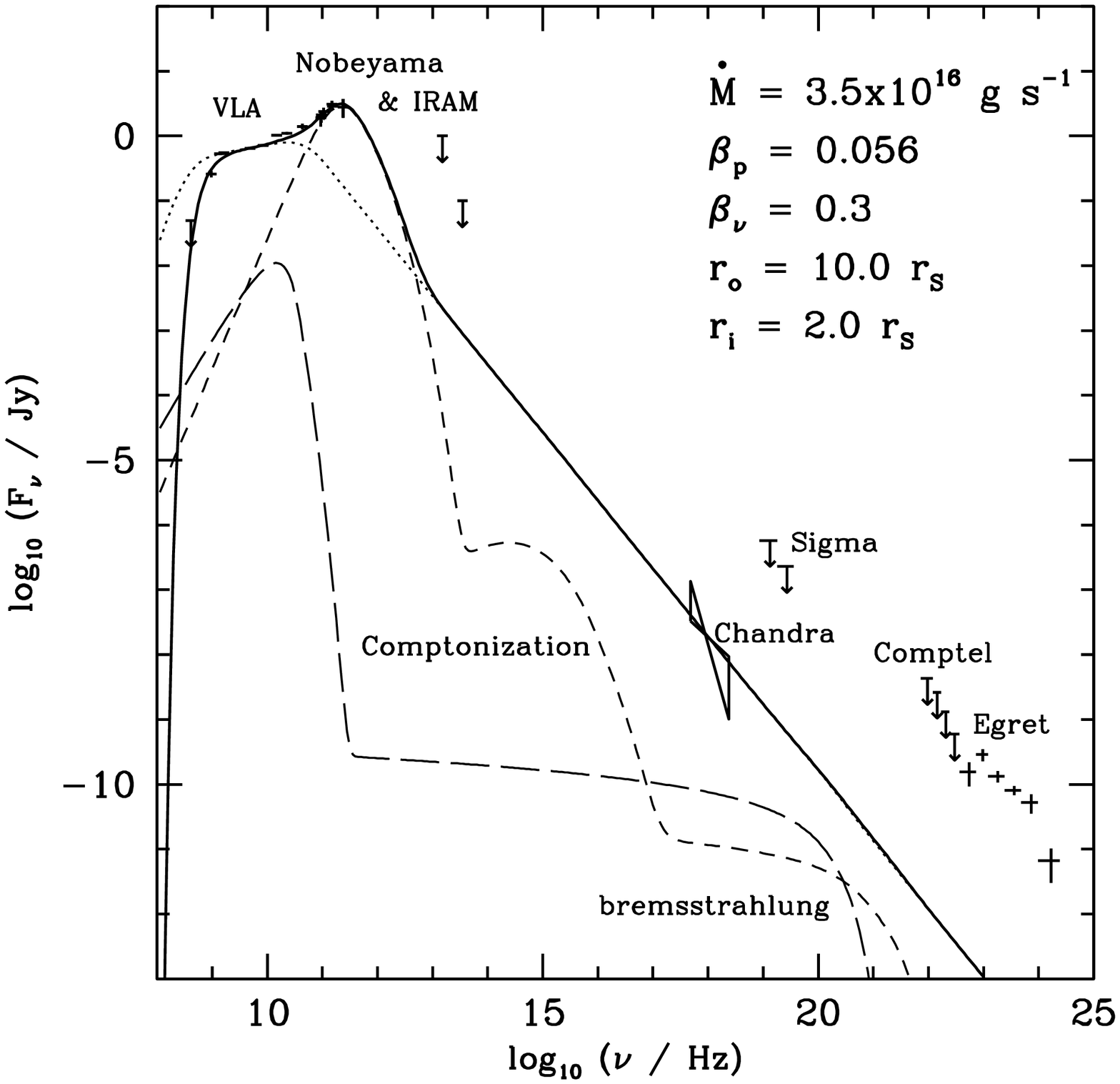}}
\end{turn}}
\caption{
Best fit spectrum for the case where the X-ray emission is dominated by 
power-law electron synchrotron radiation from the circularization region. Solid line: 
overall spectrum, corrected for free-free absorption by the ionized gas along
the line-of-sight. Dotted line: intrinsic emission from the circularization region. 
Dashed line: emission from the inner Keplerian region (see Melia et al. 2000,
2001). Long dashed line: thermal emission from the circularization region.  Here,
$r_i$ is the inner radius of the Keplerian structure.  The other parameters are
defined in the text.  The power-law electron distribution function is
$N(E,r)= 1.5\times 10^{-12} E^{-3.1}n(r)$, where $n(r)$ is the total electron 
number density. The ratio of thermal energy to the sum of gravitational energy and 
energy associated with azimuthal gas motion is $f_t=0.1$.} 
\end{figure}

\begin{figure}[thb]\label{fig2.ps}  
{\begin{turn}{0}
\epsscale{0.8}
\centerline{\plotone{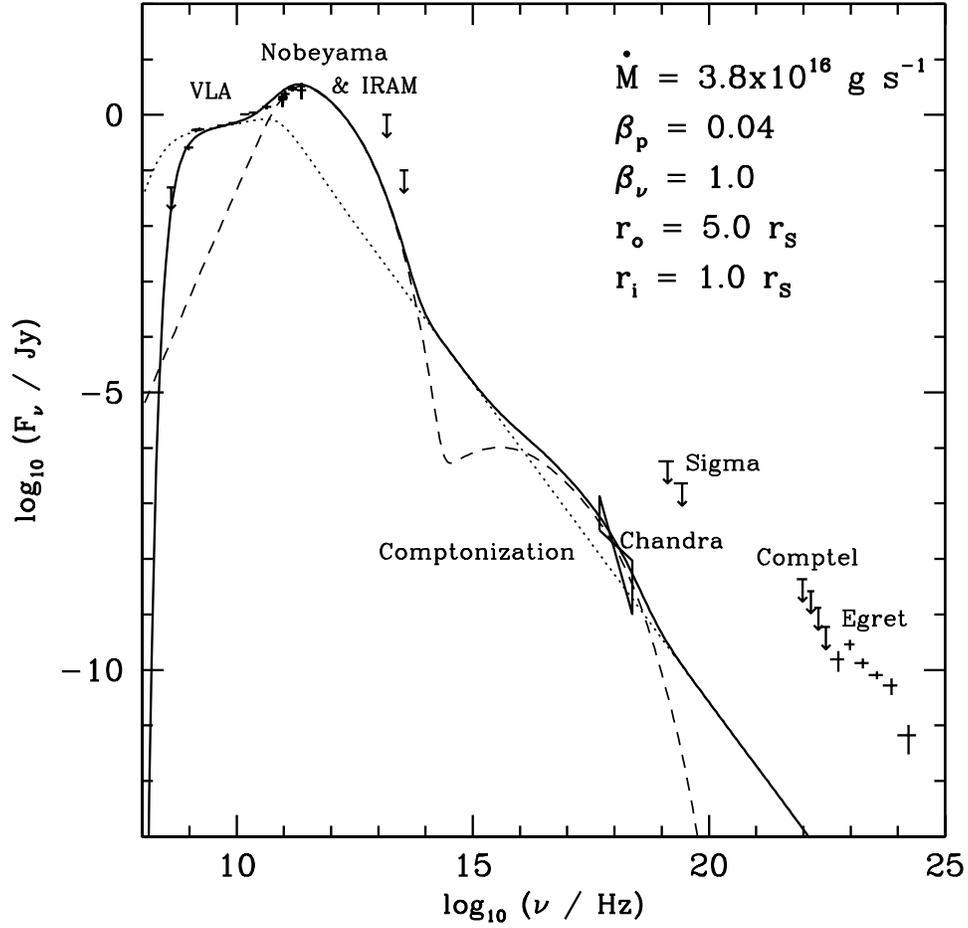}}
\end{turn}}
\caption{
Best fit spectrum for the case where the X-ray emission dominated by 
self-Comptonization of mm and sub-mm photons in the inner Keplerian 
region. The line-types are the same as those in Figure 1. Here, the 
power-law electron distribution function is $N(E,r)= 7.0\times 10^{-11} 
E^{-3.3}n(r)$ and $f_t=1.0$.}
\end{figure}

\end{document}